\documentclass[aps,prl,groupedaddress,twocolumn,10pt, superscriptaddress]{revtex4-1}

\usepackage{amsmath}
\usepackage{amssymb}
\usepackage{amsthm}
\usepackage{bbm}
\usepackage{graphicx}
\usepackage{amsmath}
\usepackage{graphicx}
\usepackage{color}
\usepackage{upgreek}
\usepackage[linkcolor = blue, citecolor = blue, urlcolor = blue, colorlinks = true]{hyperref}
\usepackage{amssymb}
\usepackage{bm}
\usepackage[capitalise]{cleveref}
\usepackage{textcomp}
\usepackage{hyperref}
\usepackage[usenames,dvipsnames]{xcolor}
\usepackage[normalem]{ulem}
\usepackage{wasysym}

\newcommand{\pers}{\, \mathrm{s^{-1}}}

\newcommand{\um}{\, \mathrm{\upmu m}}

\newcommand\diff{\mathrm{d}}
\renewcommand{\vec}[1]{\mathbf{#1}}
\renewcommand{\imath}[0]{\mathsf{i}}

\hypersetup{colorlinks=true, linkcolor=BrickRed, urlcolor=blue!50!black, citecolor=blue!50!black}

\definecolor{ABpurple}{RGB}{128, 0, 128} 
\definecolor{ABred}{RGB}{255, 0, 0} 
\definecolor{ABgreen}{RGB}{0, 255, 0} 
\definecolor{ABbrown}{RGB}{128, 64, 0} 
\definecolor{ABblue}{RGB}{0, 0, 255} 

\begin{document}
\title{{Probing the spatiotemporal dynamics of catalytic Janus particles with single-particle tracking and differential dynamic microscopy}}
\author{Christina Kurzthaler} 
\affiliation{Institut f\"ur Theoretische Physik, Universit\"at Innsbruck, Technikerstra{\ss}e 21A, A-6020 Innsbruck, Austria}
\author{Cl\'emence Devailly}
\affiliation{School of Physics and Astronomy, University of Edinburgh, James Clerk Maxwell Building, Peter Guthrie Tait Road, Edinburgh EH9 3FD, United Kingdom}
\author{Jochen Arlt}
\affiliation{School of Physics and Astronomy, University of Edinburgh, James Clerk Maxwell Building, Peter Guthrie Tait Road, Edinburgh EH9 3FD, United Kingdom}
\author{Thomas Franosch}
\affiliation{Institut f\"ur Theoretische Physik, Universit\"at Innsbruck, Technikerstra{\ss}e 21A, A-6020 Innsbruck, Austria}
\author{Wilson C. K. Poon}
\affiliation{School of Physics and Astronomy, University of Edinburgh, James Clerk Maxwell Building, Peter Guthrie Tait Road, Edinburgh EH9 3FD, United Kingdom}
\author{Vincent A. Martinez}
\email{vincent.martinez@ed.ac.uk, abrown20@staffmail.ed.ac.uk}
\affiliation{School of Physics and Astronomy, University of Edinburgh, James Clerk Maxwell Building, Peter Guthrie Tait Road, Edinburgh EH9 3FD, United Kingdom}
\author{Aidan T. Brown}
\email{vincent.martinez@ed.ac.uk, abrown20@staffmail.ed.ac.uk}
\affiliation{School of Physics and Astronomy, University of Edinburgh, James Clerk Maxwell Building, Peter Guthrie Tait Road, Edinburgh EH9 3FD, United Kingdom}
\date{\today}

\begin{abstract}
We demonstrate differential dynamic microscopy and particle tracking for the characterization of the spatiotemporal behavior of active Janus colloids in terms of the intermediate scattering function (ISF). We provide an analytical solution for the ISF of the paradigmatic active Brownian particle model and find striking agreement with experimental results from the smallest length scales, where translational {diffusion and self-propulsion dominate}, up to the largest ones, which probe effective diffusion due to rotational Brownian motion. At intermediate length scales, characteristic oscillations resolve the crossover between directed motion to orientational relaxation and allow us to discriminate active Brownian motion from other reorientation processes, e.g., run-and-tumble motion. A direct comparison to theoretical predictions reliably yields the rotational and translational diffusion coefficients of the particles, the mean and width of their speed distribution, and the temporal evolution of these parameters. 
\end{abstract}

\maketitle

Micro-swimmers exhibit a vast variety of propulsion mechanisms, ranging from turning helical screws in bacteria~\cite{Berg:1972} to the phoretic motion of synthetic, patterned (Janus) particles~\cite{Paxton:2004, Howse:2007, Jiang:2010,Lee:2014, Brown:2014}. These self-propelled agents are intrinsically out of equilibrium and display peculiar dynamical behavior due to the interplay of persistent swimming motion and stochastic fluctuations~\cite{Vicsek:2012,Romanczuk:2012,Marchetti:2013,Elgeti:2015,Bechinger:2016}. An intriguing feature displayed by micro-swimmers is the randomization of their swimming motion at large length scales due to various reorientation mechanisms, e.g., rotational diffusion expected for catalytic Janus colloids, or the `tumbling' typical of flagellated bacteria. These reorientation processes are significant for both collective dynamics~\cite{Cates:2013,Khatami:2016} and microbiology where many organisms achieve chemotaxis by varying their tumbling rate~\cite{Berg:1972,Taktikos:2014}. 

Experimental techniques for characterizing micro-swimmer behavior can be classified broadly into single-particle tracking~\cite{Howse:2007} and ensemble techniques like dynamic light scattering~\cite{Berne:1976,Lee:2014} or differential dynamic microscopy (DDM)~\cite{Cerbino:2008, Wilson:2011}. These approaches are complementary, with different merits and drawbacks. Single-particle tracking provides direct access to individual trajectories and therefore full statistical information. However, tracking requires optical resolution of single particles, and tracking of bulk (3D) systems is challenging because particles disappear from the image plane. 3D tracking is possible using Lagrangian microscopes~\cite{Berg:1972,Darnige:2017} or holographic microscopy~\cite{Sheng:06}, but these techniques are limited in statistical accuracy or limited to low particle concentration, respectively. 

In contrast, DDM provides high-throughput measurements of the intermediate scattering function (ISF) \cite{Cerbino:2008, Wilson:2011}. It is suitable for low-resolution microscopy with a large field of view, so can access large length scales, and is not restricted to low particle concentration~\cite{Poon:2016,Sentjabrskaja:2016}. DDM usually gives the ensemble-averaged dynamics over entire populations, so that using DDM to extract information on single-particle dynamics is non-ideal. For non-interacting particles, the ISF constitutes the Fourier transform of the probability density of the particle displacements and encodes full statistical information about particle behavior at a given length scale $l$ and delay time $\tau$. Hence, a quantitative comparison of experimental ISFs to theoretical predictions over a broad range of $l$ and $\tau$ should permit full characterization of the motion and extraction of key dynamical quantities.

To date, the ISF of self-propelled particles has only been measured, via DDM, at relatively small length scales for which directed motion dominates the dynamics~\cite{Wilson:2011,Lu:2012,Martinez:2012,Jepson:2013, Martinez:2014,Wittmeier:2015,Poon:2016}. Larger length scales over which reorientation leads to random motion have not been accessed yet. Theoretical predictions for the ISF of active agents have been elaborated recently probing different modes of reorientation. In particular, analytical expressions of the ISF have been obtained for the paradigmatic active Brownian particle (ABP) model in 3D~\cite{Kurzthaler:2016}, where  the orientation undergoes rotational diffusion, and for run-and-tumble particles (RTPs) in 2D~\cite{Rudnick:2004, Martens:2012}, which perform instantaneous, temporally uncorrelated tumbling events. 

In this Letter, we use DDM and particle tracking to characterize the spatiotemporal behavior of self-propelled catalytic Janus particles at a glass surface. We measure the ISF over a wide range of length and time scales and show that characteristic features emerge at intermediate length scales, where orientational relaxation occurs. These features allow us to discriminate between different propulsion models. Fitting of the ISFs using the ABP model allows extraction of key dynamical quantities, i.e. the translational ($D$) and rotational ($D_\text{rot}$) diffusion coefficients; and the mean $\langle v\rangle$ and width $\sigma_v$ of the swimming speed distribution.

{\it Experimental Method.---}Self-propelled Janus colloids~\cite{Howse:2007} were manufactured by sputter-coating ${r=0.96\pm0.04~\um}$ radius green fluorescent polystyrene colloids (Invitrogen) with a 5-nm-thick hemispherical Pt shell~\cite{Brown:2014, Brown:2016}. We suspended the Janus colloids at volume fraction 10$^{-5}$ in 0.75$\%$ w/w aqueous H$_2$O$_2$ solutions (Acros) and placed this suspension in chambers assembled from glass microscopy slides (Menzel) and $22\times 22~{\rm mm}^2$ glass coverslips (Bettering) with $\sim300~\um$ Parafilm spacers. The Janus particles self-propel by decomposing the H$_2$O$_2$ on their Pt face~\cite{Howse:2007}, though the detailed propulsion mechanism remains uncertain~\cite{Brown:2014, Ebbens:2014, Brown:2017}. The particles are bottom-heavy~\cite{Campbell:2013}, so swim towards the upper surface of the chamber, and then slide stably along that surface oriented approximately parallel to the surface in a quasi-2D layer ~\cite{Brown:2014,Dietrich:2017}. We captured a 40 min-long sequence of epifluorescence images (512$\times$512 pixels) at 50 fps (exposure time $T_e=0.02~\text{s}$), consisting of 15 sub-movies of $\approx$ 8000 images, using an inverted microscope (Ti Eclipse, Nikon) and low magnification objective (Nikon Plan Fluor 10xPh1, NA=0.3) with a sCMOS camera (Hamamatsu Orca-Flash 4.0) and Micromanager~\cite{Micromanager:2010}. Images were recorded with 4x4 pixel binning giving a $2.6~\upmu{\rm m}/$pixel resolution, at 23-25${}^\circ$C, and focusing on the upper glass surface. The number of particles per field of view increases from $\approx 60$ to $\approx 90$ particles after 15 min.

The time-sequence of images was analyzed via DDM~\cite{Wilson:2011, Martinez:2012} and standard particle tracking algorithms~\cite{Brown:2014, Brown:2016}. From particle tracking, the ISFs of the particles were computed via $f(k,\tau)=\langle e^{-\imath \vec{k}\cdot\Delta\vec{r}_j(\tau)}\rangle$, with $\Delta\vec{r}_j(\tau)$ the displacement of particle $j$ at lag time $\tau$, $\vec{k}$ the wavevector with magnitude $k=|\vec{k}|$ probing the dynamics at length scale $l=2\pi/k$, and brackets $\langle\cdot\rangle$ denoting averages over all particles $j$ and orientations of the wavevector. A fraction of tracks displaying `non-ideal' behavior (i.e. getting stuck to glass or swimming away from the glass) were discarded resulting in $\approx 60$ `well-behaved' tracks ($\gtrsim 120$~s long) of particles at the surface. For DDM analysis, the differential image correlation function, $g(\vec{k},\tau)$, i.e., the time-averaged power spectrum of the difference between pairs of images separated by time $\tau$, was computed, $g(\vec{k}, \tau)=\left\langle\left|I(\vec{k}, t+\tau)-I(\vec{k}, t)\right|^2\right\rangle_t$, with $I(\vec{k},t)$ the Fourier transform of the image $I(\vec{r},t)$. Under appropriate imaging conditions and for isotropic motion~\cite{Wilson:2011,Martinez:2012}, $g(k,\tau)=A(k)\left[1-f(k,\tau)\right]+B(k)$, with $A(k)$ the signal amplitude and $B(k)$ the camera noise. $A(k)$ is obtained from the long-time limit of $g(k,\tau)$ where $f(k,\tau)\rightarrow 0$, while in the present case of epifluorescence imaging the camera noise is negligible~\cite{Note1}.

{\it Theory.---}The motion of an ABP {in 2D} consists of isotropic translational diffusion with coefficient $D$ and directed self-propulsion at a fixed speed $v$ along the particle's instantaneous orientation $\vec{u}=(\cos\vartheta,\sin\vartheta)$, which undergoes rotational diffusion with coefficient $D_\text{rot}$. Thus, an ABP's trajectory displays a characteristic persistence length $L=v/D_\text{rot}$. We describe the ABP model {in 2D} using a Fokker-Planck equation~\cite{Sevilla:2015} (for the Langevin equations see~\cite{Note1})
\begin{eqnarray}
\partial_\tau \mathbb{P} &= -v \vec{u}\cdot\nabla_\vec{r} \mathbb{P} + D_{\text{rot}}\partial_{\vartheta}^2 \mathbb{P}+D\nabla^2_{\vec{r}}\mathbb{P}\,, \label{eq:fokkerplanck} 
\end{eqnarray}
with $\mathbb{P}(\Delta\vec{r},\vartheta,\tau|\vartheta_0)$ the probability density for an ABP to undergo a displacement $\Delta\vec{r}$ and reorient from an initial angle $\vartheta_0$ to a final angle $\vartheta$ in time~$\tau$, and $\nabla_\vec{r}$ the spatial gradient. The terms on the right-hand side correspond to propulsion, rotational, and translational diffusion, respectively. The ISF is obtained by a spatial Fourier transform, $\widetilde{\mathbb{P}}(\vec{k},\vartheta,\tau|\vartheta_0) =\int \diff^2 r \exp(-\imath\vec{k}\cdot\vec{r})\mathbb{P}(\vec{r},\vartheta, \tau|\vartheta_0)$, averaged over $\vartheta_0$ and integrated over $\vartheta$
\begin{equation}
 f(k,\tau) = \langle e^{-\imath \vec{k}\cdot\Delta\vec{r}(\tau)}\rangle = \int\!\diff\vartheta\int\!\frac{\diff\vartheta_0}{2\pi}\widetilde{\mathbb{P}}(\vec{k},\vartheta,\tau|\vartheta_0). \label{eq:ISFdef}
\end{equation}
The equation of motion for $\widetilde{\mathbb{P}}$ is solved by {following a similar solution strategy as for an ABP in 3D~\cite{Kurzthaler:2016}}, by separating variables in terms of angular eigenfunctions~\cite{Note1}. This yields an exact expression for the ISF
\begin{equation} 
	\label{finalF} f(k,\tau)\!=\! e^{-k^2D\tau}\sum_{n=0}^{\infty}\!e^{-\lambda_{2n}\tau}\!\left[\int_0^{2\pi} \!\!\frac{\diff\vartheta}{2\pi}\ \text{ce}_{2n}(q,\vartheta/2)\right]^2,
\end{equation}
where the even, $\pi$-periodic Mathieu functions $\text{ce}_{2n}(q,\vartheta)$~\cite{Ziener:2012} have imaginary deformation parameter $q= 2\imath kL$ and $\lambda_{2n}=a_{2n}(q)D_\text{rot}/4$ with $a_{2n}(q)$ the eigenvalues of the Mathieu functions.  
\begin{figure*}[t]
\centering
  \includegraphics[width=\linewidth]{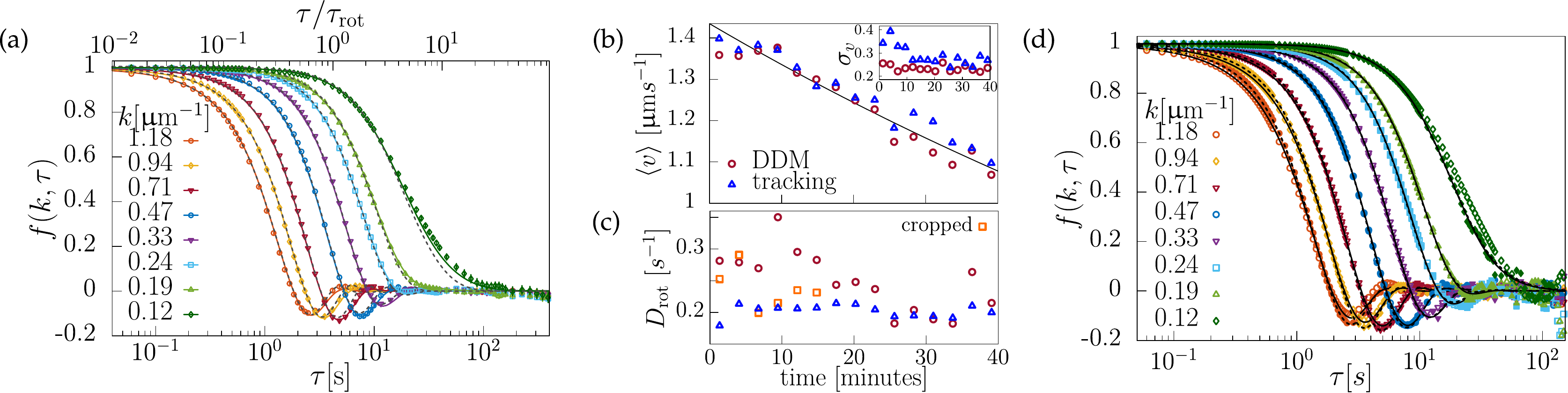}
  \caption{\label{fig_1}(a) ISFs of a dilute suspension of Janus particles obtained via DDM (symbols {and error bars}) over the whole time-sequence ($1.2\times 10^5$ images), fitted with the ABP model (coloured lines) and the RTP model (dashed, gray lines), both averaged over a Gaussian speed distribution. $\tau_\text{rot}=D_\text{rot}^{-1}$ denotes the rotational diffusion time. Global fitting of the ISFs~\cite{Note1} using the ABP model yields: {$\langle v\rangle = 1.23\pm0.01~\upmu\text{m}\text{s}^{-1}$, $\sigma_v = 0.26\pm0.02~\upmu\text{m}\text{s}^{-1}$, $D_\text{rot}=0.24\pm0.01~\text{s}^{-1}$, and $D=0.24\pm0.02~\upmu\text{m}^2\text{s}^{-1}$}. 
{Separately fitting the RTP model provides the tumbling rate $\lambda = 0.20\pm0.03~\text{s}^{-1}$, and $\langle v\rangle = 1.21\pm0.03~\upmu\text{m}\,\text{s}^{-1}$, $\sigma_v = 0.05\pm0.01~\upmu\text{m}\,\text{s}^{-1}$, $D=0.26\pm0.02~\upmu\text{m}^2\text{s}^{-1}$.} (b-c) Temporal variation of the motility parameters obtained from the ABP model fit of the ISFs extracted from DDM and tracking of fifteen $160~{\rm s}$ sub-movies. The black line is $\propto\exp{(-t/T)}$ with $T=140$~min. (d) ABP-fitted ISFs from DDM (hollow symbols, solid line) and tracking (filled symbols, dashed line) for the sub-movie at $25$~min.}
\end{figure*}

For comparison, {the motion of a RTP is characterized by straight-run phases interrupted by instantaneous tumbling events which randomize the swimming direction. The tumbling events are exponentially distributed with rate $\lambda$ (see Ref.~\cite{Martens:2012}, Eq.~(1)).} The ISF of a 2D RTP is~\cite{Martens:2012}
\begin{equation} 
\label{finalF_RTP} f(k,\tau)\!=\!e^{-k^2D\tau}\sum_{n=0}^{\infty}
\frac{e^{-\lambda\tau}\sqrt{\pi}}{\Gamma\left((n+1)/2\right)}\left(\frac{\lambda^2\tau}{2kv}\right)^{n/2}\!{\text{J}}_{n/2}(kv\tau),
\end{equation}
with ${\rm J}_{n/2}(\cdot)$ the Bessel function of order $n/2$. The ISFs for ABPs [Eq.~\eqref{finalF}] and RTPs [Eq.~\eqref{finalF_RTP}] share identical forms in the large and small $k$ regimes, where the reorientation mechanisms have not yet set in or are no longer resolved (see SI~\cite{Note1}). In particular, both ISFs can be approximated to $f(k,\tau)\approx {\rm J}_0(vk\tau)\exp(-Dk^2\tau)$ at short times, $\tau\lesssim \tau_\text{rot}:=1/D_\text{rot}$ and $\tau\lesssim 1/\lambda$ for ABPs and RTPs, respectively. At these small times and large wavenumbers $kD/v\gtrsim 1$: ${\rm J}_0(vk\tau)\to 1$ and $f(k,\tau)\approx \exp(-Dk^2\tau)$. In contrast, for intermediate wavenumbers, characteristic oscillations emerge due to the swimming motion encoded in the Bessel function. At long times and small wavenumbers $kL\ll 1$, $f(k,\tau)\approx\exp(-D_{\rm eff}k^2\tau)$ with effective diffusion $D_{\rm eff}=v^2/2D_{\rm rot} + D$ for ABPs or with $D_{\rm rot}$ replaced by $\lambda$ for RTPs. At $k$ values probing the reorientation mechanisms, $kL\simeq 2\pi$, the two ISFs are expected to display distinct behaviors reflecting rotational diffusion or random tumbling, respectively. Thus, precise measurement of the ISF at intermediate length scales should distinguish between ABPs and RTPs. By contrast, the mean-square displacements (MSD) for these two models and a range of similar models, e.g., particles simultaneously exhibiting rotational diffusion and tumbling, are identical for all lag times~\cite{Ebbens:2010,Martens:2012}.

{\it Results.---} Fig.~\ref{fig_1}(a) shows the ISF of catalytic Janus particles measured via DDM for a large window of wavenumbers and lag times. As predicted by the ABP model, the ISF exhibits oscillations at intermediate times and wavenumbers $k\gtrsim0.3~\upmu\text{m}^{-1}$, which fall off for $\tau\gtrsim4$~s due to rotational diffusion and evolve to an effective diffusive behavior for small $k\lesssim0.3~\upmu\text{m}^{-1}$; at the highest $k\approx 1.2~\upmu \text{m}^{-1}$, translational diffusion begins to damp out the oscillations. Fitting of the experimental ISFs using Eq.~\eqref{finalF} and considering a Gaussian speed distribution $P(v)$, with mean $\langle v\rangle$ and width $\sigma_v$, shows good quantitative agreement over all time and length scales considered. {The largest amplitude peak at $k=0.71~\upmu\text{m}^{-1}$ corresponds to approximately twice the persistence length, $L=\langle v\rangle /D_\text{rot}\simeq~5~\upmu\text{m}$, characterizing the transition from directed motion to effective diffusion.}

Part of $\sigma_v$ comes from particles slowing down due to H$_2$O$_2$ consumption. Analysis of short sections of the original video reveals a time-dependence in the mean propulsion speed $\langle v\rangle$, Fig.~\ref{fig_1}~(b), consistent with the exponential decay previously obtained in reaction-rate measurements on the same system~\cite{Brown:2014}. {We estimate that over 40 min this exponential decay should contribute $\approx 0.1~\upmu{\rm ms^{-1}}$ to the measured standard deviation.} The standard deviation itself would also be expected to decay exponentially, by about $20\%$, but this is smaller than the noise in this parameter (inset Fig.~\ref{fig_1}~(b)). $D$ (not shown) is constant within error, and an average over the experimental time window gives $D = 0.233\pm 0.003~\um^2\pers$. The apparent time-dependence of $D_{\rm rot}$ is due to a few bright, actively rotating particles present at short times; cropping the early videos to remove these features reduces this effect significantly, Fig.~\ref{fig_1}(c)~\cite{Note1}.

The computed ISFs from single-particle tracks show very good agreement with the ISFs obtained from DDM (see Fig.~\ref{fig_1}~(d)). {The deviations at $k=0.12~\upmu\text{m}^{-1}$ at long times, where no reliable plateau is observed, can be solely traced back to noise in the data, since the delay times here are almost equal to the length of each submovie $\sim160~\text{s}$. Parameter estimation of the submovies reveals that} the speed and the rotational diffusion coefficient (following cropping) agree reasonably between DDM and tracking (Fig.~\ref{fig_1}~(b-c)). We find $D_\text{rot}=0.226\pm 0.008~\pers$ and $0.202\pm 0.003~\pers$ from DDM and tracking, respectively. {Separately applying a `windowing' correction~\cite{Giavazzi:2017}, which aims to correct for DDM artefacts introduced by particles moving out of the field of view, yielded $D_\text{rot}=0.206\pm0.012~\text{s}^{-1}$, with other parameters left largely unchanged but more noisy~\cite{Note1}.}

By contrast, we observe a systematic deviation of the average translational diffusion coefficient between tracking ($D =0.172\pm0.002~\um^2\pers$) and DDM ($D = 0.233\pm 0.003~\um^2\pers$). This difference can partially be explained by the presence of out-of-focus particles in the bulk, {which have a higher diffusivity}. Removing some of these out-of-focus particles (by thresholding prior to DDM analysis) reduces the effect, giving $D=0.207\pm0.003~\um^2\pers$ for DDM. 

The measured $D_{\rm rot}$ agrees with the Stokes-Einstein prediction for equivalent equilibrium particles in the bulk, $D_\text{rot}^\text{SE}=k_\text{B}T/(8\pi\eta r^3)=0.21\pm0.03~\text{s}^{-1}$ with solution viscosity $\eta=0.91\pm0.02 ~{\rm mPa\,s}$ estimated from literature values~\cite{Phibbs:1951, Haynes:2013}. However, $D$ is lower than the bulk free diffusion coefficient of an equivalent passive colloid $D^\text{SE}=k_BT/(6\pi\eta r)=0.25\pm~0.02\um^2\pers$. Proximity to the wall will likely account for some of this difference, just as for a passive colloid~\cite{Happel:2012}, {for which one indeed expects only translational diffusion to be significantly perturbed~\cite{Leach:2009}.}

Unlike the MSD, the ISF should allow us to discriminate between rotational diffusion and run-and-tumble motion. However, we see in Fig.~\ref{fig_1}(a) that the RTP model also reproduces surprisingly well the main features of the experimental ISFs. This is because of the qualitative similarity between the ABP and RTP ISFs, and because the finite width of the speed distribution, $P(v)$, inherently present in the swimmer population, gives an additional degree of freedom through the fitting parameter, $\sigma_v$. {We also separately fitted each submovie, which removes the temporal part of the speed distribution (for $160~{\rm s}$ segments, the expected temporal contribution is only $7~{\rm nm\,s^{-1}}$), enhancing the distinction between the swimming mechanisms. Significantly better agreement with the ABP model is obtained for all submovies: the mean-squared error (MSE), for the ABP model is $(1.02\pm0.09)\times 10^{-4}$, whereas for the RTP model, it is roughly $50\%$ larger at $(1.53\pm0.08)\times 10^{-4}$.}

\begin{figure}
\includegraphics[width=\linewidth]{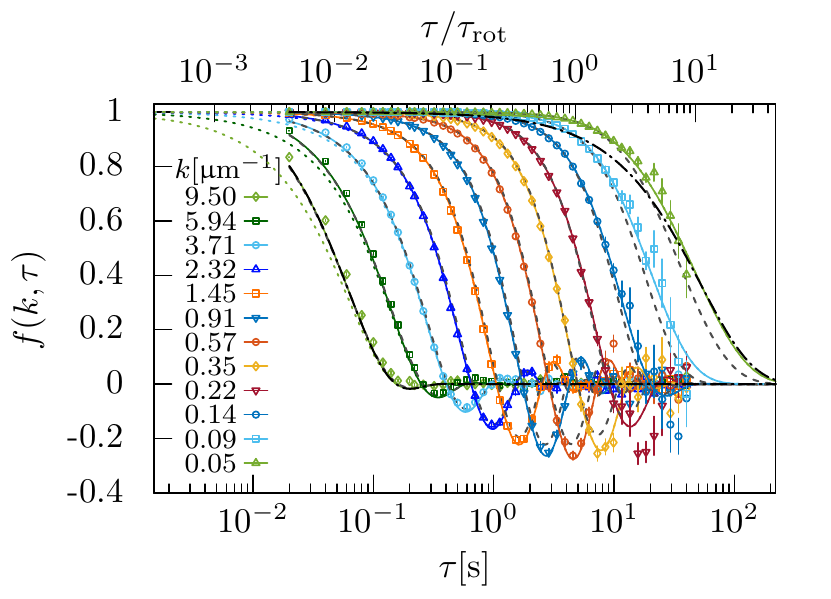}
  \caption{\label{fig_3}ISFs for a single Janus particle (symbols {and error bars, as in Fig.~\ref{fig_1}}). Coloured solid lines correspond to the theoretical predictions for ABPs ({$v=1.64\pm0.02 ~\upmu \text{m/s}$, $D=0.16\pm0.01~\upmu \text{m}^2\text{/s}$, and $D_{\text{rot}}=0.21\pm0.02~ \text{s}^{-1}$)} and gray dashed lines for equivalent RTPs {($v = 1.58\pm0.02~\upmu \text{m/s}$, $D = 0.17\pm 0.01~\upmu \text{m}^2\text{/s}$, and $\lambda = 0.14\pm0.01~\text{s}^{-1}$)}, both corrected for a finite exposure time $T_e$. Coloured dashed lines correspond to the uncorrected ABP ISF [Eq.~\eqref{finalF}]. Dash-dotted, black lines indicate the corrected large and uncorrected small wavenumber approximations, $\exp[-k^2D(\tau-T_e/3)]\text{J}_0(vk\tau)$ and $\exp(-k^2D_\text{eff}\tau)$, respectively. Note that the ISF does not approach the small wavenumber approximation closely even at the smallest $k$ displayed, for which $kL=0.43$.}
\end{figure}

To strengthen these results, we investigated the ISF computed from single-particle tracks, e.g., Fig.~\ref{fig_3}, recorded at higher magnification (x20 Nikon Plan Apo Ph1, NA=0.75, 1x1 binning, 351 nm/pixel) to give access to higher $k$ values. Fitting the ISFs with the ABP and RTP models (now with a single speed $v$) reveals small deviations at short times ($\tau<0.01\tau_\text{rot}$) for $k\gtrsim~2.3\upmu \text{m}^{-1}$, but correcting for the effect of the finite exposure time $T_e$ removes these deviations (Fig.~\ref{fig_3} and SI~\cite{Note1}). 

As expected, both ABP and RTP models agree at short times where the swimming direction has not changed due to rotational diffusion or instantaneous tumbling events. For $\tau\lesssim\tau_\text{rot}$ and large $k$, oscillations are strongly damped by translational diffusion, tending towards a single exponential decay. Deviations between the ABP and RTP models become apparent for $\tau\simeq\tau_\text{rot}$ and $kL\simeq2\pi$ (in Fig.~\ref{fig_3}: $k\simeq0.80~\upmu\text{m}^{-1}$), where the underlying reorientation mechanisms determine the particle dynamics (plotting the fitting residuals highlights these deviations, see SI~\cite{Note1}). These results confirm the systematic small quantitative difference between the RTP model and the experimental data observed in DDM, while the ABP model closely reproduces the experimental ISFs over 2 and 3 decades of length- and time-scales, respectively. We repeated this analysis with, in total, 23 `well-behaved' single-particle trajectories; in all but 4 of these, the ABP model reproduced the data better, {with MSE $=(4.43\pm 1.03)\times 10^{-4}$, than the RTP model, MSE $=(5.46\pm0.65)\times10^{-4}$ (averaged over all 23 trajectories, and for $\tau\lesssim 10\text{ s}$)}.

{\it Summary and Conclusion.---}We have characterized the spatiotemporal dynamics of hydrogen-peroxide-fueled Janus colloids using DDM and particle tracking. Experimental observations of the ISF showed striking agreement with theoretical predictions of the active Brownian particle model over a broad range of length and time scales, reflecting the transition from directed swimming motion to the randomization of the orientation. The ISFs allowed us to distinguish between different modes of orientational relaxation (continuous rotational diffusion versus instantaneous tumbling) by probing the dynamics directly at the relevant length scale, i.e., the persistence length of the active agents. Additionally, we have demonstrated DDM as a high-throughput method to extract relevant motility parameters (and their temporal variation) of ensembles of catalytic Janus colloids.

Most previous studies of active colloids focused on the MSD obtained from single-particle tracking~\cite{Howse:2007,Jiang:2010, Ebbens:2012, Brown:2014, Ebbens:2014, Brown:2016, Ebbens:2010, Dietrich:2017} which does not discriminate between ABPs, RTPs, or similar models, in contrast to the ISF. One prior study measured higher moments of the particle motion, e.g., the non-Gaussian parameter, which should differentiate between different types of orientational relaxation~\cite{Zheng:2013}, but comparison with the ABP model yielded no quantitative agreement there. This might result from statistical uncertainties or from, e.g., anisotropic translational diffusion~\cite{Kurzthaler:2016}. 

Here, we showed that the standard ABP model fully describes the dynamics of phoretically-driven Janus colloids even at small length scales $l$ that go down to a fraction of the particle diameter, i.e. $l/2r\approx0.35$ (for $k=9.5~\upmu m^{-1}$). This indicates that microscopic details of the propulsion mechanism~\cite{Golestanian:2005,Popescu:2016, Golestanian:2007, Brown:2014, Brown:2017} can be coarse-grained into a few minimal, mesoscopic processes: propulsion, rotational diffusion, and translational diffusion, discussed here, as well as anisotropic diffusion~\cite{Kurzthaler:2016} and deterministic rotational drift~\cite{Kurzthaler:2017:circle}, which are relevant for anisotropic or chiral Janus particles~\cite{Ebbens:2010,Kummel:2013, tenHagen:2014} and swimming bacteria near walls~\cite{Lauga:2006,diLeonardo:2011}.

{In particular, our results show that extensions to the ABP model to include, e.g., temporal variations in particle speed~\cite{Romanczuk:2011} are not necessary. Such effects have been hypothesized to arise from hydrodynamic and other interactions between the active particle and the surface leading to e.g., spontaneous oscillations~\cite{Lintuvuori:2016, Shen:2018} and thermal fluctuations in particle orientation with respect to the surface~\cite{Zheng:2013}. Though our results show that these effects do not significantly perturb the ABP behavior within our experimental window, probing these effects at shorter time and/or length scales remains an interesting research avenue.} 

We anticipate that our results will serve as a reference for the dynamics of synthetic self-propelled particles irrespective of their propulsion mechanism. Moreover, DDM provides a powerful tool to probe the dynamics of micro-swimmers over large length scales, and overcomes the statistical limitations of 3D single-particle tracking and poor optical resolution. Therefore, it can be efficiently applied to investigate the spatiotemporal dynamics of active particles moving in 3D~\cite{Jepson:2013, Martinez:2014, Palacci:2010} or dense suspensions ~\cite{Theurkauff:2012,Palacci:2013,Paoluzzi:2014}, or of smaller size ~\cite{Lee:2014,Ma:2015}. In particular, DDM might shed light on the self-propulsion of enzyme-based nanomotors~\cite{Sengupta:2013,RiedelC:2015} in the presence of strong stochastic forces, the run-and-tumble behavior of biological micro-swimmers, such as {\it E. coli} bacteria~\cite{Berg:1972}, or the chemotactic response of synthetic vesicles, which have potential for drug delivery~\cite{Joseph:2017}.

The research data presented in this publication is available on the Edinburgh DataShare repository~\cite{note2}.


\begin{acknowledgments}
{\it Acknowledgments} We thank Benno Liebchen and Sebastian Leitmann for helpful and stimulating discussions. This work was supported by the Austrian Science Fund (FWF): P 28687-N27; the UK EPSRC: grant EP/J007404/1; 
and the ERC: Advanced Grant ERC-2013-AdG 340877-PHYSAP.
\end{acknowledgments}


%

\end{document}




\title{{Probing the spatiotemporal dynamics of catalytic Janus particles with single-particle tracking and differential dynamic microscopy: Supplementary Information}}
\author{Christina Kurzthaler} 
\affiliation{Institut f\"ur Theoretische Physik, Universit\"at Innsbruck, Technikerstra{\ss}e 21A, A-6020 Innsbruck, Austria}
\author{Cl\'emence Devailly}
\affiliation{School of Physics and Astronomy, University of Edinburgh, James Clerk Maxwell Building, Peter Guthrie Tait Road, Edinburgh EH9 3FD, United Kingdom}
\author{Jochen Arlt}
\affiliation{School of Physics and Astronomy, University of Edinburgh, James Clerk Maxwell Building, Peter Guthrie Tait Road, Edinburgh EH9 3FD, United Kingdom}
\author{Thomas Franosch}
\affiliation{Institut f\"ur Theoretische Physik, Universit\"at Innsbruck, Technikerstra{\ss}e 21A, A-6020 Innsbruck, Austria}
\author{Wilson C. K. Poon}
\affiliation{School of Physics and Astronomy, University of Edinburgh, James Clerk Maxwell Building, Peter Guthrie Tait Road, Edinburgh EH9 3FD, United Kingdom}
\author{Vincent A. Martinez}
\email{vincent.martinez@ed.ac.uk, abrown20@staffmail.ed.ac.uk}
\affiliation{School of Physics and Astronomy, University of Edinburgh, James Clerk Maxwell Building, Peter Guthrie Tait Road, Edinburgh EH9 3FD, United Kingdom}
\author{Aidan T. Brown}
\email{vincent.martinez@ed.ac.uk, abrown20@staffmail.ed.ac.uk}
\affiliation{School of Physics and Astronomy, University of Edinburgh, James Clerk Maxwell Building, Peter Guthrie Tait Road, Edinburgh EH9 3FD, United Kingdom}
\date{\today}

\maketitle
   
\section{Intermediate Scattering Function of an Active Brownian Particle in 2D}
\subsection{Analytic solution}
In its simplest form, the motion of an active Brownian particle (ABP) in a plane consists of isotropic translational diffusion with diffusivity $D$, and directed self-propulsion at constant velocity $v$ along the particle's instanteneous orientation $\vec{u}=(\cos\vartheta,\sin\vartheta)$. The orientation performs rotational diffusion characterized by the rotational diffusion coefficient $D_\text{rot}$. Hence, the Langevin equations for the position $\vec{r}$ and the orientation $\vartheta$
of the ABP assume the form~\cite{Ebbens:2010,tenHagen:2011,Sevilla:2015}
\begin{subequations}
 \label{fokkerplanck}
\begin{align}
\label{trans}\diff \vec{r}(\tau) 	&= v\vec{u}(\tau)\diff \tau + \sqrt{2D} \boldsymbol{\xi}(\tau)\diff \tau,\\
\diff \vartheta(\tau)	&= \sqrt{2D_\text{rot}}\zeta(\tau) \diff \tau,
\end{align}
\end{subequations}
where $\boldsymbol{\xi}(\tau)$ and $\zeta(\tau)$ are independent Gaussian white noise
processes with zero mean and variance $\langle \xi_i(\tau)\xi_j(\tau')\rangle =
\delta_{ij}\delta(\tau-\tau')$ for $i,j=1,2$ and
$\langle\zeta(\tau)\zeta(\tau')\rangle=\delta(\tau-\tau')$. 
Using standard techniques~\cite{Gardiner:2009}, we obtain the Fokker-Planck equation for the probability density
$\mathbb{P}\equiv\mathbb{P}(\Delta\vec{r},\vartheta,\tau|\vartheta_0)$ of a particle to
displace a distance $\Delta\vec{r}$ and to reorient from an initial orientation
$\vartheta_0$ to a final orientation $\vartheta$ in a lag time~$\tau$, 
\begin{align}
\partial_\tau \mathbb{P} &= -v \vec{u}\cdot\nabla_\vec{r} \mathbb{P} + D_{\text{rot}}\partial_{\vartheta}^2 \mathbb{P}+D\nabla^2_{\vec{r}}\mathbb{P}, \label{eq:fokkerplanck} 
\end{align}
where $\nabla_\vec{r}$ denotes the spatial gradient. The corresponding initial condition is
$\mathbb{P}(\Delta\vec{r},\vartheta,\tau=0|\vartheta_0) = \delta(\Delta\vec{r})\delta(\vartheta-\vartheta_0 \text{ mod }2\pi)$. 
Here, we provide an exact solution for the intermediate scattering function (ISF),
\begin{align}
  f(\vec{k},\tau)  	&=\langle e^{-\imath \vec{k}\cdot\Delta\vec{r}(\tau)}\rangle
			= \int_0^{2\pi}\diff\vartheta\int_0^{2\pi}\frac{\diff\vartheta_0}{2\pi} \ \widetilde{\mathbb{P}}(\vec{k},\vartheta,\tau|\vartheta_0),\label{eq:ISFdef}
\end{align}
which is obtained by averaging over the initial orientation $\vartheta_0$ and integrating over the final orientations $\vartheta$ of the spatial 
Fourier transform of the probability density, 
\begin{align}
\widetilde{\mathbb{P}}(\vec{k},\vartheta,\tau|\vartheta_0)  &=
\int_{\mathbb{R}^2} \diff^2 r
\exp(-\imath\vec{k}\cdot\vec{r})\mathbb{P}(\vec{r},\vartheta, \tau|\vartheta_0),
\end{align}
with wavevector $\vec{k}$. To solve the associated equation of motion,
\begin{align}
 \partial_\tau\widetilde{\mathbb{P}}&= -\imath v\vec{u}\cdot\vec{k}\widetilde{\mathbb{P}}+D_\text{rot}\partial^2_\vartheta\widetilde{\mathbb{P}}-D\vec{k}^2\widetilde{\mathbb{P}},\label{eq:fptransform}
\end{align} 
{we follow a similar solution strategy as employed in our previous work~\cite{Kurzthaler:2016,Kurzthaler:2017,Kurzthaler:2017:circle}}. In particular,
we choose the wavevector in direction, $\vec{k}=k\vec{e}_1$ and use a separation of variables in terms of 
appropriate angular eigenfunctions, $\exp(-\lambda t)z(\vartheta)${, to elaborate the exact solution for the ISF}. 
{Thus, Eq.~\eqref{eq:fptransform} simplifies to a non-Hermitian Sturm-Liouville problem for the angular eigenfunctions~$z(\vartheta)$ 
\begin{align}
\left[D_\text{rot}\frac{\diff^2}{\diff \vartheta^2} - \imath v k \cos\vartheta  -k^2D+\lambda\right]z(\vartheta)&=0, 
\end{align}
where $\lambda$ denotes the separation constant.}
A change of variable, $x=\vartheta/2$, leads to the eigenvalue problem
\begin{align}
 \left[\frac{\diff^2}{\diff x^2}+\bigl(a-2q\cos (2x)\bigr)\right]z(x)&=0,\label{eq:Mathieu}
\end{align}
which has the form of the Mathieu equation~\cite{NIST:online,NIST:print,Ziener:2012} with (imaginary) deformation
parameter {$q= 2\imath vk/D_\text{rot}=2\imath k L$, where $L$ denotes the persistence length,} and 
eigenvalue {$a=4(\lambda-k^2D)/D_\text{rot}$}. 
Hence, the angular eigenfunctions $z(x)$ are $\pi$-periodic even and odd Mathieu functions, {$\text{ce}_{2n}(q,x)$ and $\text{se}_{2n+2}(q,x)$,}
with associated eigenvalues {$a_{2n}(q)$} and {$b_{2n+2}(q)$}, respectively.
{The Mathieu functions constitute a complete, orthogonal, and normalized set of eigenfunctions in the sense of 
$\int_0^{2\pi}\!\diff x \ \text{ce}_{2n}(q,x)\text{ce}_{2m}(q,x)=\pi\delta_{nm}$
and similarly for $\text{se}_{2n+2}(q,x)$~\cite{NIST:online,NIST:print,Ziener:2012}. Moreover, they can be expressed in terms of deformed cosines and sines: 
\begin{align}
\text{ce}_{2n}(q,x)&=\sum_{m=0}^\infty A_{2m}^{2n}(q)\cos\left(2mx\right),\label{eq:even}\\
\text{se}_{2n+2}(q,x)&= \sum_{m=0}^\infty B_{2m+2}^{2n+2}(q) \sin\left[(2m+2)x\right].\label{eq:odd} 
\end{align}
By inserting these expansions [Eq.~\eqref{eq:even}-\eqref{eq:odd}] into the Mathieu equation [Eq.~\eqref{eq:Mathieu}], the recurrence
relations for the Fourier coefficients of the even Mathieu functions, $A_{2m}^{2n}(q)$, are obtained
\begin{align}
  a_{2n}A_0^{2n}-qA_2^{2n}  &=0,\label{eq:rec1}\\
  (a_{2n}-4)A_2^{2n}-q(2A_0^{2n}+A_4^{2n})  &=0,\\
(a_{2n}-4m^2)A_{2m}^{2n}-q(A_{2m-2}^{2n}+A_{2m+2}^{2n})&=0 \ \ \ \text{ for } m\geq 2,\label{eq:rec2}
\end{align}
and similar relations hold for the Fourier coefficients of the odd 
Mathieu functions, $B_{2m+2}^{2n+2}(q)$.
}

Then, the analytic solution of the Fourier transform $\widetilde{\mathbb{P}}$ is expressed as a superposition of these eigenfunctions~\cite{NIST:online,NIST:print}
{
\begin{align}
 \widetilde{\mathbb{P}}(\vec{k},x,\tau|x_0)&= \sum_{n=0}^\infty\left[\alpha_{2n}\text{ce}_{2n}(q,x)e^{-\lambda_n\tau}+\beta_{2n+2}\text{se}_{2n+2}(q,x)e^{-\lambda_n\tau}\right],\label{eq:expansion}
\end{align}
where the coefficients ($\alpha_{2n}$, $\beta_{2n+2}$) of the expansion are determined such that the initial condition for $\tau=0$ is fulfilled:
\begin{align}
 \widetilde{\mathbb{P}}(\vec{k},x,\tau=0|x_0) = \frac{1}{2}\delta(x-x_0 \text{ mod } \pi) = \sum_{n=0}^\infty\left[\alpha_{2n}\text{ce}_{2n}(q,x)+\beta_{2n+2}\text{se}_{2n+2}(q,x)\right].
\end{align}
In particular, multiplying both sides by $\text{ce}_{2m}(q,x)$ (or $\text{se}_{2m+2}(q,x)$), integrating over $x$, and using the 
orthogonality relations, yields the coefficients of the expansion:
\begin{align}
\alpha_{2n} &= \frac{1}{2\pi} \int_0^{2\pi}\!\diff x \ \delta\!\left(x-x_0\right)\text{ce}_{2n}(q,x) = \frac{1}{2\pi}\text{ce}_{2n}(q,x_0);\\
\beta_{2n+2} &= \frac{1}{2\pi} \int_0^{2\pi}\!\diff x \ \delta\!\left(x-x_0\right)\text{se}_{2n+2}(q,x) = \frac{1}{2\pi}\text{se}_{2n+2}(q,x_0).
\end{align}  
Collecting our results provides the full solution for the Fourier transform:}
\begin{align}
\widetilde{\mathbb{P}}(\vec{k},\vartheta,\tau|\vartheta_0) \!=\! \frac{e^{-k^2D\tau}}{2\pi}\sum_{n=0}^\infty\left[
\text{ce}_{2n}(q,\vartheta_0/2)\text{ce}_{2n}(q,\vartheta/2)e^{-D_\text{rot}a_{2n}(q)\tau/4}
  +\text{se}_{2n+2}(q,\vartheta_0/2)\text{se}_{2n+2}(q,\vartheta/2)e^{-D_\text{rot}b_{2n+2}(q)\tau/4}\right].
\end{align}
 
After integration [Eq.~\eqref{eq:ISFdef}] the exact expression of the ISF reduces to
\begin{align} 
  f(k,\tau)	&= e^{-k^2D\tau}\sum_{n=0}^{\infty}e^{-D_\text{rot}a_{2n}(q)\tau/4}\left[\int_0^{2\pi}\frac{\diff\vartheta}{2\pi} \ \text{ce}_{2n}(q,\vartheta/2)\right]^2\,,\label{eq:ISF}
\end{align}
where the integrals over the odd Mathieu functions vanish by symmetry.

{Finally, we discuss the uniqueness of the solution and outline important properties of the 
Mathieu functions.\\ 
\textbf{Uniqueness of the solution.}
We show that Eq.~\eqref{eq:fptransform} has at most one solution and, thus, $\widetilde{\mathbb{P}}$ is a unique solution. 
Therefore, we first assume that a second solution $\widetilde{\mathbb{P}}_0$
exists and consider the difference $\omega=\widetilde{\mathbb{P}}-\widetilde{\mathbb{P}}_0$. The equation of motion for $\omega$ is the same as Eq.~\eqref{eq:fptransform}, 
yet with a different initial condition $\omega(\vec{k},\vartheta,\tau=0|\vartheta_0)=0$. 
We consider the integral over all angles $\vartheta$, $\Omega(\tau)=\int\!\diff\vartheta \ |\omega|^2$, 
and take the derivative with respect to time $\diff \Omega(\tau)/\diff \tau = \int\!  \diff\vartheta \ \left(\omega^*\partial_\tau\omega+\omega\partial_\tau\omega^*\right)$, 
where $\omega^*$ denotes the complex conjugate of $\omega$ with equation of motion 
$\partial_\tau\omega^*=\imath v\vec{u}\cdot\vec{k}\omega^*+D_\text{rot}\partial^2_\vartheta\omega^*-D\vec{k}^2\omega^*$.
Inserting the equations of motion for $\omega$ and $\omega^*$, and as $\omega$ and $\omega^*$ vanish at the boundaries due to the 
$2\pi$-periodicity in the angle $\vartheta$, 
we arrive at 
\begin{align}
 \frac{\diff \Omega(\tau)}{\diff \tau} &=  -2Dk^2 \int_0^{2\pi}\! \diff\vartheta \ |\omega|^2 - 2D_\text{rot}\int_0^{2\pi}\!\diff\vartheta \ \left|\partial_\vartheta \omega\right|^2\leq 0.   
\end{align}
Thus, $\Omega(\tau)$ does not increase for all times $\tau$. As  $\Omega(\tau=0)=0$ and $\Omega(\tau)\geq 0$ $\forall \tau$,
$\omega=\widetilde{\mathbb{P}}-\widetilde{\mathbb{P}}_0$ has to vanish 
for all times $\tau$ and, therefore, the solution $\widetilde{\mathbb{P}}$ is unique.\\

\textbf{Properties of Mathieu functions and eigenvalues. } 
The eigenfunctions and eigenvalues solve a non-Hermitian eigenvalue problem [Eq.~\eqref{eq:Mathieu}]  
and are therefore in general complex. It has been elaborated in Ref.~\cite{Ziener:2012} that the 
eigenvalues of the Mathieu functions are real for small (purely imaginary) deformation parameters (here: $q=2\imath kL$), 
whereas they branch out to two complex conjugates at increasing $|q|$. In particular, at  $|q|=2kL\simeq1.47$ the lowest real eigenvalues, 
$a_0(q)$ and $a_2(q)$, merge and become complex conjugates. The same occurs for higher eigenvalues at even larger $|q|$~\cite{Ziener:2012}.
As complex eigenvalues always appear in complex-conjugated pairs, the solution of the ISF [Eq.~\eqref{eq:ISF}] remains a real function.  
Interestingly, the first branching point sets the length scale for the appearance of oscillations in the ISF (see Figs. 1a,d and 2 in the main text), 
as has been found earlier for the ABP model in 3D~\cite{Kurzthaler:2016}. 
Moreover, it determines the maximal radius of convergence for a low-wavenumber approximation of the ISF.}

\subsection{Numerical evaluation}
Interestingly, the integral over the even Mathieu functions in
Eq.~\eqref{eq:ISF} reduces to the zeroth Fourier coefficient of the Mathieu function
{$\int_0^{2\pi}\diff\vartheta \ \text{ce}_{2n}(q,\vartheta/2)/2\pi= A_0^{2n}(q)$}~\cite{NIST:online,
NIST:print,Ziener:2012}, and the ISF reduces to
\begin{align}
  f(k,\tau) &= e^{-k^2D\tau}\sum_{n=0}^{\infty}\left[A_0^{2n}(q)\right]^2e^{-D_{\text{rot}}a_{2n}(q)\tau/4}. \label{eq:ISF_numerics}
\end{align}
{Since the eigenvalues of the Mathieu functions are ordered with increasing 
real part $\text{Re}[a_0(q)]\leq\text{Re}[a_2(q)]\leq\ldots$ higher terms in the expansion are exponentially suppressed  
and the series rapidly converges.} 
The Fourier coefficients $A_0^{2n}(q)$ can be evaluated numerically
by {solving the matrix eigenvalue problem associated with the 
recurrence relations [Eq.~\eqref{eq:rec1}-\eqref{eq:rec2}]}~\cite{Ziener:2012}: 
\begin{align}
 \begin{pmatrix}
  0 & \sqrt{2}q &  & & 0\\
\sqrt{2}q & 4 & q & &\\
& q & \ddots & \ddots & \\
  & & \ddots & \ddots  & q \\
  0 &  & & q & 4M^2
 \end{pmatrix}
\begin{pmatrix}
\sqrt{2}A_{0}^{2n}\\
A_{2}^{2n}\\
\vdots \\
\vdots \\
A_{2M}^{2n}
\end{pmatrix} = 
a_{2n}\begin{pmatrix}
\sqrt{2} A_{0}^{2n}\\
 A_{2}^{2n}\\
 \vdots \\
\vdots \\
 A_{2M}^{2n}
\end{pmatrix},
\end{align}
truncated at an appropriate dimension $M$  such that the normalization condition at $\tau=0$, $\sum_n [A_0^{2n}(q)]^2=1$,
is fulfilled with desired accuracy. {Throughout we used $M = 100$ to solve the eigenvalue problem and truncated the series in Eq.~\eqref{eq:ISF_numerics} after 40 terms.}

\subsection{Persistent motion and effective diffusion}
For large wavenumbers and for times $\tau\lesssim \tau_\text{rot}$ where rotational diffusion (or tumbling $\tau\lesssim 1/\lambda$) 
has not yet set in, the particle moves persistently along a constant orientation $\vec{u}_0=(\cos\vartheta_0,\sin\vartheta_0)$ and 
is subject to translational diffusion. The corresponding ISF evaluates to 
\begin{align}
 f(k,\tau)	&= e^{-k^2D\tau}\int_0^{2\pi}\frac{\diff\vartheta_0}{2\pi} e^{-\imath v k \cos\vartheta_0 \tau} = e^{-k^2D\tau}J_0(vk\tau),
\end{align}
where $J_0(\cdot)$ denotes the Bessel function of order zero. 

{For long times the ISF is dominated by the first term of the expansion [Eq.~\eqref{eq:ISF_numerics}], as the eigenvalues
are ordered with increasing real part,
\begin{align}
 f(k,\tau) &\simeq e^{-Dk^2\tau}\left[A_0^0(q)\right]^2e^{-D_\text{rot}a_0(q)\tau/4} \quad \text{ for } \tau\to \infty. \label{eq:ISF_numerics_ap}
\end{align}
In particular, for small deformation parameters $|q|\ll 1.47$ (i.e. $kL \ll 0.735$) the zeroth eigenvalue of the Mathieu functions can be 
approximated by $a_0(q) = |q|^2/2+\mathcal{O}(|q|)^4$ and the zeroth Fourier coefficient by $A_0^0(q)=1+\mathcal{O}(|q|)^2$~\cite{Ziener:2012}.

Inserting these expressions into Eq.~\eqref{eq:ISF_numerics_ap} yields 
\begin{align}
 f(k,\tau)&\simeq\exp(-k^2D_\text{eff}\tau),
\end{align}
with effective diffusion coefficient $D_\text{eff}=D+v^2/2D_\text{rot}$, which agrees with the slope of the mean-square displacement for 
long times~\cite{Ebbens:2010,Sevilla:2015}. Thus, at large length-scales the particle undergoes enhanced effective diffusion. 

More generally, the low-order moments can be extracted from the ISF by a perturbative approach in the wavenumber $k$, 
see Refs.~\cite{Kurzthaler:2016,Kurzthaler:2017:circle} for an ABP in 3D and a Brownian circle swimmer in 2D.} 

\section{Data Analysis}
\subsection{Analysis of the intermediate scattering function obtained via differential dynamic microscopy (DDM) and single-particle tracking}

We obtain the ISF from DDM via Eq.(2) of the main text. We estimate $A(k)$ as the mean of $g(k,\tau)$ for 
{$37~{\rm s}<\tau<123~{\rm s}$}.
{Since the differential image correlation function $g(k,\tau)$ for $k=0.12~\upmu\text{m}^{-1}$ decorrelates at times larger than $37~\text{s}$, 
we obtain $A(k)$ as the mean of $g(k,\tau)$ for $128~{\rm s}<\tau<430~{\rm s}$ in Fig. 1a of the main text and 
$76~{\rm s}<\tau<154~{\rm s}$ in Fig. 1d of the main text, respectively.}

Furthermore, we set $B(k)=0$ because in fluorescence microscopy the camera noise 
is usually negligible. We checked this assumption by {fitting $B(k)$ rather than setting it to zero. Therefore, we estimated} 
$B(k)$ from a linear extrapolation of $g(k,\tau)$ to $\tau=0$ for each $k$, and found that the ratio $A(k)/B(k)$ ranges 
from $10^{-4}$ to $10^{-2}$. {Including the estimated $B(k)$ in the fitting procedure gave no 
significant difference for the motility parameters than from using $B(k)=0$, and, thus, confirms that the camera noise $B(k)$ is indeed negligible}. 
{On the contrary, non-fluorescence imaging (for example bright-field or phase contrast imaging) is expected to produce a larger, 
non-negligible camera noise $B(k)$, which can be obtained from the plateau of $g(k,\tau)$ 
at short times. For high wavenumbers, the plateau might only be present at shorter times and thus movies should be collected at a faster frame rate. 
However, $B(k)$ could still be estimated as an additional parameter in the 
fitting procedure.}

In our experiments a few large, bright particles displayed an anomolous swimming behavior containing periods of strong, active rotational motion. 
We believe this generates an anomolously high fitted $D_{\rm rot}$ because active rotation also generates decorrelation of the 
swimming direction. To remove these particles we cropped the videos: a region of 256$\times$256 pixels was selected from the 
original 512$\times$512 pixels video specifically to avoid including the anomolous particles. This reduced the apparent $D_{\rm rot}$, 
as shown in Fig. 1c of the main text.

For the particle tracking accompanying the DDM analysis we excluded all trajectories not on the glass surface by 
using an intensity threshold. We also excluded all trajectories of duration $<120~{\rm s}$ and any trajectories showing 
neglible motion over continuous periods of $10{~\rm s}$ or greater, which we defined as stuck particles.

\textbf{Estimation of motility parameters.}
From both DDM and tracking data we fitted the ISF, $f(k,\tau)$, to several wavenumbers, 
($k\in [0.24,0.48,0.71,$ $0.94,1.18]~\upmu \text{m}^{-1}$) {for lag times $\tau\leq 36~\text{s}$} simultaneously in order to capture the different 
time- and length-scales of the system. Fitting employed nonlinear least-square minimization based on the bound-constrained 
BFGS (Broyden-Fletcher-Goldfarb-Shanno) algorithm in Python~\cite{Byrd:1995} yielded the motility parameters 
$v,\sigma_v, D_\text{rot}$, and $D$. We assume the speed $v$ of the Janus particles to be normally distributed with 
mean $\langle v \rangle$ and standard deviation $\sigma_v$ and obtain the ISF by numerically post-averaging the 
analytical result in Eq.~\eqref{eq:ISF}. Note that negative $v$ do not contribute significantly as the standard 
deviation is much smaller than the mean velocity. 

The ISF, $f(k,\tau)$, of a single particle (Fig. 2 in the main text), which moves at constant velocity $v$, 
was fitted by using data for several wavenumbers $k\in[0.2,6.0]\upmu\text{m}^{-1}$
and lag times $\tau\leq18\text{s}$ simultaneously. At large wavenumbers and short times artefacts due to
exposure time become relevant and need to be accounted for (for details see Sec.~\ref{sec:exposure_time}).  

{
To obtain an error estimate for the fitted parameters $\hat{p} \in \{\langle v\rangle,\sigma_v,D, D_\text{rot}\}$ 
(or $\hat{p} \in \{v,D, D_\text{rot}\}$) we applied a standard jackknife 
resampling method~\cite{efron:1981}. In our global fits, we usually estimate the motility parameters $\hat{p}$ by fitting datasets of $N$ different 
wavenumbers $\{k_1,\hdots,k_N\}$ simultaneously. Jackknife resampling relies on estimating the parameters $p_i$ from the data, where the information of 
wavenumber $i$ has been omitted and only data for wavenumbers $\{k_1,\hdots,k_{i-1},k_{i+1},\hdots,k_N\}$ has been fitted. 
This procedure is repeated for all wavenumbers. The standard jackknife error (SE) is then obtained by  
\begin{align}
 \text{SE}(\hat{p})	&= \left[\frac{N-1}{N}\sum_{i=1}^N \left(\bar{p}-p_i\right)^2\right]^{1/2},
\end{align}
where $\bar{p}=\sum_{i=1}^Np_i/N$ denotes the mean of the parameter estimates $p_i$ obtained by omitting wavenumber~$i$.
}

\begin{figure}[htp]
\centering
\includegraphics[width = 0.7\linewidth]{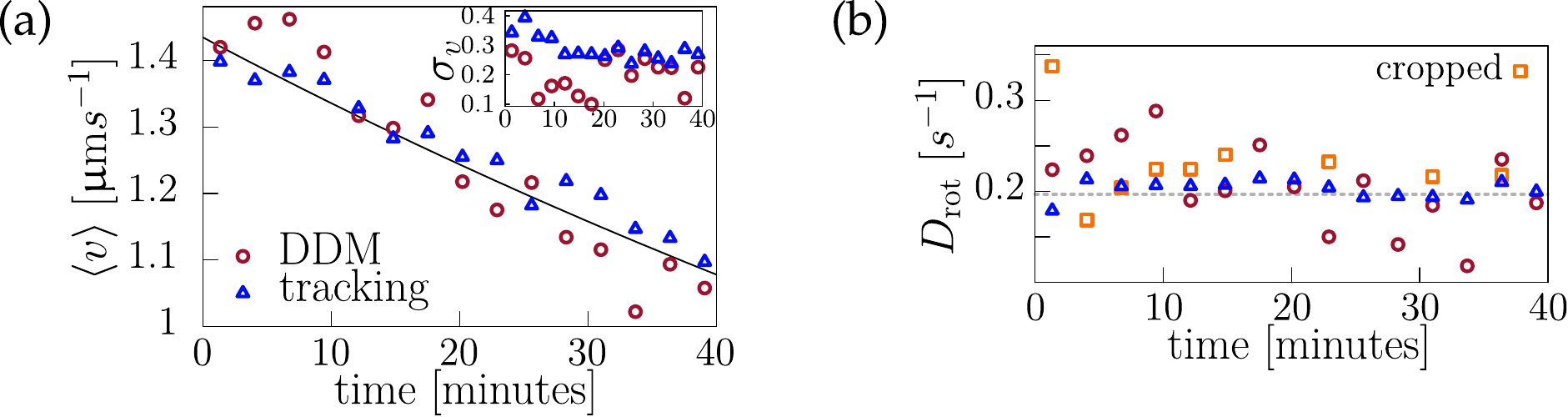}
\caption{{Motility parameters (a) $\langle v\rangle$ and $\sigma_v$ (inset), and (b) $D_\text{rot}$ obtained for each submovie by using window analysis prior to DDM processing. 
On average the motility parameters evaluate to $\langle v \rangle = 1.249\pm 0.037~\upmu \text{m}\pers$, $\sigma_v = 0.200\pm0.016~\upmu \text{m}\pers$, 
$D_\text{rot}=0.206\pm0.012\pers$, and $D=0.239\pm0.003~\um^2\pers$. For comparison, the motility parameters 
obtained from standard image processing yielded: $\langle v \rangle =1.237 \pm 0.028~\upmu \text{m}\pers$, $\sigma_v = 0.235\pm0.003~\upmu \text{m}\pers$, $D_\text{rot}= 0.226 \pm 0.008\pers$ (including cropping), and $D=0.233\pm 0.003~\um^2\pers$. 
Their temporal variation is shown in Figs.1b-c in the main text.}} 
\label{fig_window}
\end{figure}

{\subsection{Image window analysis}}
{It is well known that finite image size leads to spectral leakage for Fourier transforms, and particles leaving or entering the field of view can thus introduce artefacts in the DDM analysis. Artefacts are effectively suppressed by applying window functions which attenuate the edges of the field of view. We follow the procedure outlined in Ref.~\cite{Giavazzi:2017} and applied a Blackman-Harris window to our movies before applying the DDM analysis. Results for the windowed analysis are shown in Fig.~\ref{fig_window}. Overall the results are very similar, within error bars, to the non-windowed analysis shown in Figs. 1b-c of the main manuscript although somewhat noisier as the windowing attenuates the overall signal. However, we found a slightly lower $D_\text{rot}=0.206\pm 0.008~\pers$ closer to the value obtained from tracking, $D_\text{rot}=0.202\pm 0.003~\pers$.}

\subsection{Correction of the intermediate scattering function for a finite exposure time~\label{sec:exposure_time}}
For small delay times the finite exposure time $T_e$ of the camera causes motion blur and must be corrected for. This has been done previously for the MSD~\cite{Savin:2005}, but to our knowledge an equivalent correction for the ISF has not been derived. Let us assume that a single frame centred at time $t$ captures all photons within time $[t-T_e/2, t+T_e/2]$, and that the particle position recorded within that time corresponds to the time-average centre of mass 
\begin{align}
\bar{\vec{r}}(t; T_e)\,=\,T_e^{-1}\int_{t-T_e/2}^{t+T_e/2} \vec{r}(t'){\rm d}t' \,,
\end{align}
where the overbar indicates recorded values as opposed to the true values, which have no overbar. Then the recorded ISF will be
\begin{align}
\bar{f}(k,\tau;T_e)\,=\,\langle e^{-\imath \vec{k}\cdot\left[\bar{\vec{r}}(\tau+t; T_e)-\bar{\vec{r}}(t; T_e)\right]}\rangle\,.
\end{align}
We can solve this equation analytically in the case of pure diffusion with diffusivity $D$. Without loss of generality, let us set $t=0$ and examine a single trajectory with initial and final frames centred at $0$ and $\tau$. The exposure time is generally shorter than the inverse frame rate, so we assume $\tau\geq T_e$. Then for a purely diffusive process, the motion within the first frame will be entirely uncorrelated with the motion within the final frame, and both will be uncorrelated with the motion at intermediate times. In other words, we can write
\begin{align}
\bar{\vec{r}}(\tau; T_e)-\bar{\vec{r}}(0)\,=\,\left[\bar{\vec{r}}(\tau; T_e)-\vec{r}(\tau-T_e/2)\right]+\left[\vec{r}(\tau-T_e/2)-\vec{r}(T_e/2)\right]+\left[\vec{r}(T_e/2)-\bar{\vec{r}}(0; T_e)\right]\,,
\end{align}
where the three terms in brackets correspond to displacements contained entirely within three non-overlapping time windows, $[-T_e/2,T_e/2]$, $[T_e/2, \tau-T_e/2]$ and $[\tau-T_e/2, \tau+T_e/2]$, corresponding respectively to the first frame, the intermediate time and the final frame. For a diffusive process, displacements within non-overlapping time windows are uncorrelated, so we can rewrite the recorded ISF as
\begin{align}
\bar{f}(k,\tau;T_e)\,=\,\langle e^{-\imath \vec{k}\cdot\left[\bar{\vec{r}}(\tau; T_e)-\vec{r}(\tau-T_e/2)\right]}\rangle \langle e^{-\imath \vec{k}\cdot\left[\vec{r}(\tau-T_e/2)-\vec{r}(T_e/2)\right]}\rangle \langle e^{-\imath\vec{k}\cdot\left[\vec{r}(T_e/2)-\bar{\vec{r}}(0; T_e)\right]}\rangle\,.
\end{align}
The first and third factors are equal by time-reversal symmetry, and the second factor is just a true ISF, so after some rearrangement
\begin{align}
\bar{f}(k,\tau;T_e)\,=\,f(k,\tau-T_e)\left\langle e^{-\imath\vec{k}\cdot\left[\bar{\vec{r}}(T_e/2; T_e)-\vec{r}(0)\right]}\right\rangle^2\,. \label{rearranged}
\end{align}
For a diffusive process, the first term is just $f(k,\tau-T_e)=e^{-Dk^2(\tau-T_e)}$. We calculate the second term by writing it out explicitly
\begin{align}
\langle e^{-\imath\vec{k}\cdot\left[\bar{\vec{r}}(T_e/2; T_e)-\vec{r}(0)\right]}\rangle = \left\langle \exp{\left\{-\frac{\imath\vec{k}}{T_e}\cdot\int_0^{T_e}\left[\vec{r}(t')-\vec{r}(0)\right] {\rm d}t'\right\}}\right\rangle\,. \label{sq term}
\end{align}
and splitting the displacement into a series of infinitessimal steps
\begin{align}
\vec{r}(t')-\vec{r}(0)=\lim_{\delta t\rightarrow 0}\sum_{j=1}^{t'/\delta t}\delta\vec{r}_j\,,
\end{align}
where $\delta t$ is a small timestep, and the small displacement $\delta\vec{r}_j=\vec{r}(j\delta t)-\vec{r}([j-1]\delta t)$. Inserting this expression into Eq.~\eqref{sq term} and interchanging the order of integration and summation gives
\begin{align}
\langle e^{-\imath\vec{k}\cdot\left[\bar{\vec{r}}(T_e/2; T_e)-\vec{r}(0)\right]}\rangle \,=\,     \left\langle \exp{\left[-\imath\vec{k}\cdot\lim_{\delta t\rightarrow 0}\sum_{j=0}^{T_e/\delta t}\left(1-\frac{j\delta t}{T_e}\right)\delta\vec{r}_j\right]}\right\rangle        \label{sq term summed}
\end{align}
where the factor in brackets $()$ comes from the fact that each infinitesimal displacement $\delta r_j$ contributes to the total displacement at all times $[j\delta t,T_e]$, i,e., for a duration $T_e-j\delta t$. Here again, for a diffusive process each $\delta r_j$ is an independent displacement, so we can take the summation outside the average as a product
\begin{align}
\langle e^{-\imath\vec{k}\cdot\left[\bar{\vec{r}}(T_e/2; T_e)-\vec{r}(0)\right]}\rangle \,=\,    \lim_{\delta t\rightarrow 0}\prod_{j=0}^{T_e/\delta t} \left\langle e^{-\imath\vec{k}\cdot\left(1-\frac{j\delta t}{T_e}\right)\delta\vec{r}_j}\right\rangle\,.        \label{product}
\end{align}
Each of the terms in the product is equivalent to an ISF over the time interval $\delta t$ with a renormalized $\vec{k}$, i.e.,
\begin{align}
\left\langle e^{-\imath\vec{k}\cdot\left(1-\frac{j\delta t}{T_e}\right)\delta\vec{r}_j}\right\rangle=f\left(k[1-j\delta t/T_e],\delta t\right)=e^{-Dk^2(1-j\delta t/T_e)^2\delta t}\,,
\end{align}
which gives
\begin{align}
\langle e^{-\imath\vec{k}\cdot\left[\bar{\vec{r}}(T_e/2; T_e)-\vec{r}(0)\right]}\rangle \,=\,   \exp{\left[-Dk^2\delta t \lim_{\delta t\rightarrow 0}\sum_{j=0}^{T_e/\delta t} (1-j\delta t/T_e)^2\right]}\,.
\end{align}
Replacing the sum with an integral and evaluating gives
\begin{align}
\langle e^{-\imath\vec{k}\cdot\left[\bar{\vec{r}}(T_e/2; T_e)-\vec{r}(0)\right]}\rangle \,=\,   \exp{\left[-Dk^2\int_0^{T_e}(1-t'/T_e)^2{\rm d}t'\right]}=e^{-Dk^2T_e/3} \label{correction}\,,
\end{align}
and inserting Eq.~\eqref{correction} into Eq.~\eqref{rearranged} yields finally 
\begin{align}
\bar{f}(k,\tau;T_e)\,=\,e^{-Dk^2(\tau-T_e/3)}\,=\,f(k,\tau)e^{Dk^2T_e/3}\,. \label{final}
\end{align}
for $\tau\geq T_e$, and $f(k,0,T_e)=1$. The calculation can also be done for $0<\tau<T_e$, but as mentioned above, this will not apply unless the frames overlap. 

It turns out that a pure self-propulsive component can be included relatively trivially because motion blur does not modify the measured trajectory for straight-line motion and this motion decouples from the diffusive motion given above. Mathematically, self-propulsion in direction $\vec{u}$ just introduces a factor $e^{-\imath \vec{k}\cdot \vec{u}v\tau}$ to the ISF. Including this factor and performing the average over $\vec{u}$ does not modify any of the calculations to obtain Eq.~\eqref{final}. Hence we have
\begin{align}
\bar{f}(k,\tau;T_e)={\rm J}_0(kv\tau)e^{-Dk^2(\tau-{T_{e}}/3)}=f(k,\tau)e^{Dk^2T_e/3} \,. \label{propulsion final}
\end{align}
Provided there is a large separation of timescales $T_e\ll \tau_{\rm rot}$, this expression also provides a good approximation for the case with rotational decorrelation, so in the main text we fit the measured ISF with
\begin{align}
\bar{f}(k,\tau; T_e)\simeq  e^{Dk^2{T_{e}}/3}f(k,\tau)\,,
\end{align}
where $f(k,\tau)$ here corresponds to the predicted ISF from the ABP or RTP models.

\subsection{Comparison of the active Brownian particle and run-and-tumble particle models}
\begin{figure*}[htp]
 \includegraphics[width = \linewidth]{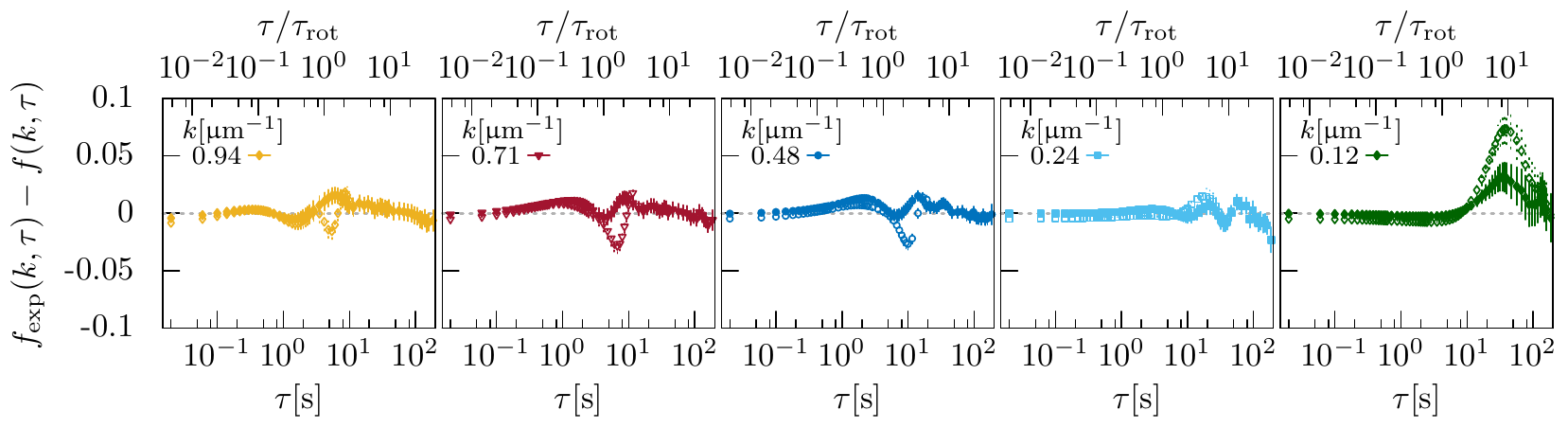}
\caption{Differences of the experimental ISF $f_\text{exp}(k,\tau)$ obtained via DDM (corresponding to Fig.1a in the main text) with respect to the ISF $f(k,\tau)$ of the ABP (bold symbols) and the RTP model (hollow symbols) for wavenumbers $k$
probing the reorientation mechanisms. Here, $\tau_\text{rot}$ denotes the rotational diffusion time. The error bars correspond to the error bars of the experimental ISF.\label{fig:residuals_ddm}}
 \includegraphics[width = \linewidth]{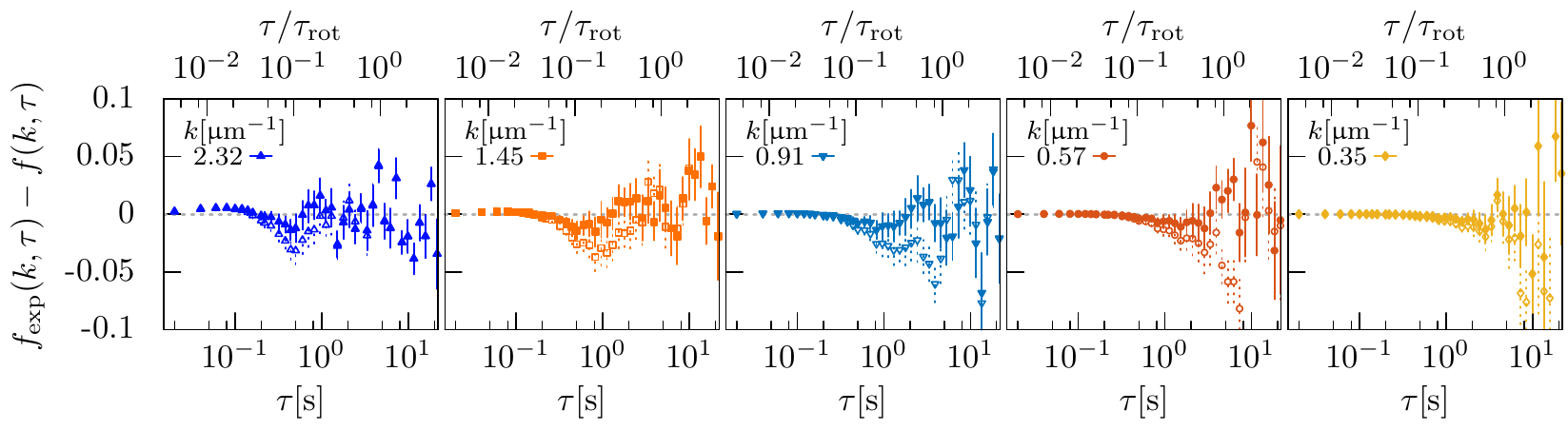}
\caption{Differences of the experimental ISF $f_\text{exp}(k,\tau)$ obtained via tracking of a single particle (corresponding to Fig.2 in the main text) with respect to the ISF $f(k,\tau)$ 
of the ABP (bold symbols) and the RTP model (hollow symbols) for wavenumbers $k$ probing the reorientation mechanisms. Here, $\tau_\text{rot}$ 
denotes the rotational diffusion time. The error bars correspond to the error bars of the experimental ISF.\label{fig:residuals}}
\end{figure*}

Here, we compare the experimental ISF obtained via DDM for a dilute suspension of Janus colloids (see Fig.1a in the main text) and the ISF of a single Janus colloid accessed via particle tracking (see Fig.2 in the main text) with the theoretical predictions for the ABP and RTP models. Therefore, we consider the residuals $f_\text{exp}(k,\tau)-f(k,\tau)$ at intermediate wavenumbers $kL\simeq2\pi$ which probe the relaxation of the orientation, see Figs.~\ref{fig:residuals_ddm} and~\ref{fig:residuals} for data obtained via DDM and single-particle tracking, respectively.

The differences between the experimental data obtained via DDM and the RTP model become most pronounced at the smallest wavenumber considered ($k=0.12~\upmu\text{m}^{-1}$), whereas at larger wavenumbers the differences between the ABP model and RTP model remain similar (Fig.~\ref{fig:residuals_ddm}). Yet, as already depicted in Fig.2 in the main text, the differences between the data of a single particle and the RTP model are significantly larger than for the ABP model (Fig.~\ref{fig:residuals}). The differences for the RTP model increase at intermediate times $\tau\simeq1/\lambda$ where the reorientation due to tumbling has already set in and the run-and-tumble dynamics differ from active Brownian motion. The residuals for the ABP model increase only at large times  due to statistical errors.